\documentclass{aa}  
\usepackage{enumitem}
\usepackage{graphicx}
\usepackage{cancel}
\usepackage{txfonts}
\usepackage[pdfencoding=auto,psdextra]{hyperref}
\usepackage{pifont}
\usepackage[export]{adjustbox}
\usepackage[dvipsnames]{xcolor}

\usepackage[dvipsnames]{xcolor}
\hypersetup{
    colorlinks=true,
    linkcolor=BrickRed,
    anchorcolor=BrickRed,
    filecolor=Violet,      
    urlcolor=Mulberry,
    citecolor=MidnightBlue,
}

\begin{document}

   \title{The AIDA-TNG project: 3D halo shapes}

   \author{C. Giocoli,
          \inst{1,2}\fnmsep\thanks{\url{carlo.giocoli@inaf.it}}
          \and
          G. Despali\inst{3,1,2}
          \and
          L. Moscardini\inst{3,1,2}
          \and
          M. Meneghetti\inst{1,2} 
          \and
          R. K. Sheth\inst{4}
          \and
          A. Pillepich\inst{5}
          \and
          M. Vogelsberger\inst{6}
          }

   \institute{
     INAF-Osservatorio di Astrofisica e Scienza dello Spazio di Bologna, Via Piero Gobetti 93/3, I-40129 Bologna, Italy
     \and
     INFN -- Sezione di Bologna, viale Berti Pichat 6/2, I-40127 Bologna, Italy
     \and
     Dipartimento di Fisica e Astronomia "Augusto Righi", Alma Mater Studiorum Università di Bologna, via Gobetti 93/2, I-40129 Bologna, Italy        
     \and
     Center for Particle Cosmology, University of Pennsylvania, Philadelphia, PA 19104, USA
     \and 
      Max-Planck-Institut für Astronomie, Königstuhl 17, D-69117 Heidelberg, Germany     
     \and
     Department of Physics, Kavli Institute for Astrophysics and Space Research, Massachusetts Institute of Technology, Cambridge, MA 02139, USA
    }

   \date{Received \textit{\color{purple}today}; Accepted \textit{\color{purple}tomorrow}: {\color{purple}\underline{we wish all referee reports were this fast!}} Published \textit{\color{purple}when the Universe allows}.}

  \abstract {The shapes of dark matter halos can be used to constrain the fundamental properties of dark matter. In standard Cold Dark Matter (CDM) cosmologies, halos are typically triaxial, with a preference for prolate configurations; however, including the full baryonic physics tends to make them more oblate.}
   {We focus on the characterization of total matter 3D shape in alternative dark matter models, such as Self-Interacting Dark Matter (SIDM) and Warm Dark Matter (WDM). These scenarios predict different structural properties due to collisional effects or the suppression of small-scale power.}
   {We measure the different halo component shapes -- dark matter, stars, and gas -- at various radii from the center in the AIDA-TNG (Alternative Interacting Dark Matter and Astrophysics – TNG), which is a suite of high-resolution cosmological simulations built upon the IllustrisTNG framework. The intent is to systematically study how
     different dark matter models -- specifically, SIDM and WDM -- affect galaxy formation and the structure of dark matter halos, when realistic baryonic physics is also included.}
   {SIDM models tend to produce rounder and more isotropic halos, especially in the inner regions, as a result of momentum exchange between dark matter particles. Group and cluster-size WDM halos are also slightly more spherical than their CDM counterparts. In all cases, the inclusion of self-consistent baryonic physics makes the central regions of all halos rounder, while still revealing clear distinctions among the various dark matter models, mainly valid for the self-interacting ones.}
   {The general framework presented in this work,
     based on the 3D halo shape, can be useful to interpret multi-wavelength data analyses of galaxies and clusters.}

   \maketitle
   %
   
\section{Introduction}

The $\Lambda$ Cold Dark Matter ($\Lambda$CDM) model has become the standard paradigm for cosmic structure formation, successfully describing the hierarchical assembly of galaxies and clusters within dark matter (DM) halos \citep{white78,bond91,tormen97a,tormen98a}. 
In this framework, DM halos arise from the nonlinear growth of primordial density fluctuations \citep{press74,bardeen86,sheth99b,musso21}, observed in the Cosmic Microwave Background \citep{planck18}, and provide the gravitational potential wells in which baryons condense to form galaxies \citep{kauffmann93,white96a,bullock01b,springel05b,baugh06}. 
DM constitutes roughly 85\% of the total matter density, while baryons contribute the remaining 15\%. 

Despite its success on large scales, $\Lambda$CDM exhibits small-scale tensions, including discrepancies with respect to observational data in central halo densities, shapes, satellite distributions, and velocity anisotropies \citep{vogelsberger12,weinberg15,vogelsberger16,bullock17,zavala19,meneghetti20}.  
These tensions have motivated the exploration of alternative dark matter scenarios, such as Self-Interacting Dark Matter (SIDM) and Warm Dark Matter (WDM). SIDM introduces non-negligible particle scattering, which can isotropize orbits, reduce halo triaxiality, and produce cored central density profiles \citep{rocha13,zavala13,robertson18,vogelsberger19, oneil23}. WDM suppresses small-scale power and delays halo formation, leading to less concentrated, more spherical halos compared to CDM \citep{lovell12,menci12}. 
In hierarchical models, the shape and structure of a DM halo can be affected by its assembly history \citep{jing02,allgood06,rossi11,despali13,bonamigo15,musso23,nikakhtar25}. Therefore, understanding the effects of alternative DM models on halo assembly and morphology is critical for interpreting both simulations and observational data \citep{dave01,viel05b,viel12,viel13,kaplinghat16,tulim18}.

Hydrodynamical simulations demonstrate that baryonic physics significantly alters halo and subhalo internal structure \citep{despali17b}. The condensation of baryons into the central region of a halo increases its sphericity and oblateness, while stellar and AGN feedback can partially counteract these effects by redistributing gas and energy \citep{gnedin04,bryan13,pillepich18,chua19}. Halo shapes vary with radius: they are typically more triaxial in the inner regions and rounder near the virial radius \citep{despali13,despali14,despali17}. Velocity anisotropies of DM particles and halo stars also depend on radius, being nearly isotropic at small radii, increasingly radial at intermediate distances, and sometimes returning to isotropy near the virial boundary \citep{wojtak05,ascasibar08}. Baryons reduce radial motions, increase tangential orbit fractions, and raise velocity dispersions, particularly in central regions.

Substructures and anisotropic accretion further influence halo morphology \citep{nipoti18}. Massive satellites often preserve their infall directions along filaments, while smaller satellites are accreted from a broader range of angles, affecting halo triaxiality in the outskirts \citep{knebe04,libeskind13a,libeskind13b,forero-romero14}. The combined effects of baryons, substructure, and filamentary accretion produce complex, radially dependent halo shapes. Observational studies infer triaxial halo shapes through gravitational lensing, X-ray measurements, the Sunyaev-Zel'dovich effect, and the spatial distribution of satellites \citep{defilippis05,oguri10,sereno13,gonzalez21}. Typically, the observable is a projected two-dimensional shape \citep{meneghetti07b}; we defer a detailed study of the inversion of this projection into 2D detectable quantities to a separate paper.

In this work, we use the AIDA-TNG simulations \citep{despali25}, a suite of high-resolution cosmological magnetohydrodynamical runs, to explore halos' three-dimensional shapes and internal properties across CDM, SIDM, and WDM scenarios. The AIDA-TNG simulations incorporate full baryonic physics, including gas cooling, star formation, and feedback processes, allowing us to quantify the interplay between baryons and DM microphysics \citep{vogelsberger14a,vogelsberger14b,rodriguez-gomez16,rodriguez-gomez17,pillepich18,vogelsberger20}. The reference DM-only runs will also allow us to quantify the effects of baryonic physics implementations and possible impact on DM particle properties.
We focus on mass-selected halo samples across different DM models, without attempting direct halo matching, to provide a statistically representative assessment of how halo morphology responds to baryonic processes and alternative DM properties.

Specifically, we characterize three-dimensional halo shapes using axis ratio parameters and investigate the radial dependence of these properties, depending on the velocity anisotropies of both DM particles, gas, and halo stars. We examine the influence of substructure and halo mass on these trends. 
By comparing CDM, SIDM, and WDM runs, we connect simulation predictions to observable signatures, providing insights relevant for ongoing and upcoming surveys such as Euclid \citep{euclidredbook,scaramella22,euclidoverview}, DESI\footnote{\url{https://datalab.noirlab.edu/data/desi}} \citep{desi_collaboration}, and JWST\footnote{\url{https://science.nasa.gov/mission/webb/}} \citep{wst_whitep}. This study thus advances our understanding of the combined impact of baryons and dark matter physics on halo structure, with implications for both theoretical and observational aspects.

The structure of this paper is as follows. In Sect. \ref{sec_sim}, we describe the cosmological simulations employed in this study. 
In Sect. \ref{sec_globalshapes}, we present the method used to measure axis ratios and discuss our results as a function of radius, considering the various dark matter models. 
In Sect. \ref{sec_misal}, we examine the misalignment of the various components, highlighting how it depends on dark matter particle models.
In Sects. \ref{sec_massdep} and \ref{sec_correlation} we study the correlation between the major-to-minor axis ratio as a function of the host halo mass and the intermediate-to-minor axis ratio.
We summarize and conclude in Sect. \ref{sec_sum}.

\section{AIDA-TNG cosmological hydrodynamical simulations}
\label{sec_sim}
The AIDA-TNG simulations, \emph{Alternative Dark Matter in the TNG universe} (AIDA for short, hereafter) \citep{despali25b,despali25,romanello25}, consist of three cosmological volumes of increasing resolution and decreasing size simulated in cold and alternative dark matter (ADM, hereafter) models, with and without baryonic physics. This work utilizes the largest boxes, measuring 75 $h^{-1}$ Mpc (110.7 Mpc) on a side (see Table \ref{table:1} for more details). The uniqueness of the AIDA set lies in the fact that we can utilize the same cosmological volumes across multiple dark matter scenarios, encompassing cold, warm, and self-interacting dark matter. In particular, we make use of the WDM3 run, corresponding to a particle mass of 3 keV, and two different self-interacting models, namely SIDM1 with a constant cross section equal to 1 cm$^2$ g$^{-1}$, and vSIDM with a scale-dependent cross section as described in \citet{correa21}.
Moreover, we can rely on both a dark matter-only (DMO) and a full-physics (FP) version of each run so that we can disentangle the effects, relative to a standard CDM run, of ADM and baryonic physics as described by the IllustrisTNG galaxy formation model \citep[TNG hereafter][]{weinberger17,pillepich18}. 

All simulations start at redshift $z=127$ and adopt the cosmological model from \citet{planckxxiv}: $\Omega_{\rm m}=0.3089$, $\Omega_{\rm \Lambda}=0.6911$, $\Omega_{\rm b}=0.0486$, $H_0=67.74\;\mathrm{km\,s^{-1}Mpc}$  and $\sigma_{8}=0.8159$. The initial conditions (ICs) have been generated with \textsc{N-GenIC} code \citep{springel05a} by applying the Zel’dovich approximation on a glass distribution of particles with a linear matter transfer function computed using the \textsc{CAMB} code \citep{camb}. The simulations were run from the same ICs as the corresponding runs of the original IllustrisTNG runs \citep{pillepich18b}. 

{\renewcommand{\arraystretch}{1.3}
\begin{table*}
\caption{The AIDA-TNG runs used in this work.}
\label{table:1}
\centering
\begin{tabular}{c c c c c c c|  c | c c }
\hline\hline
Name & Box & Physics & $m_{\rm DM}$ & $m_{\rm baryons}$ & $\epsilon_{\rm DM,*}^{z=0}$ & CDM & WDM3  & SIDM1  & vSIDM \\
& $h^{-1}\rm{[Mpc]}$&  & $ [h^{-1}\rm{M}_{\odot}$] & $ [h^{-1}\rm{M}_{\odot}$] & [$h^{-1}$kpc] & & 3\,keV & 1$\,\rm{cm^{2}g^{-1}}$ & \citet{correa21} \\
\hline\hline        
   100/A& 75 & dark-matter only & 4.80$\times 10^{7}$ & - & 1 &\ding{51}  &\ding{51}  &\ding{51}  &\ding{51} \\
   & & full-hydro & 4.04$\times 10^{7}$ & 7.55$\times 10^{6}$ & 1 & \ding{51} & \ding{51} & \ding{51}  & \ding{51} \\
\hline\hline
\end{tabular}
\end{table*}
}

Halos and subhalos are identified using a two–step procedure that combines a Friends-of-Friends (FOF) algorithm with the \texttt{SUBFIND} substructure finder \citep{davis85, springel01b, dolag09}. In the first step, the FOF algorithm groups dark matter particles into halos using a standard linking length of $b = 0.2$ times the mean inter–particle separation, corresponding approximately to an overdensity of $\sim 200$ times the mean cosmic density. This step provides a catalog of virialized structures, each potentially containing multiple gravitationally bound subhalos. In the second step, the \texttt{SUBFIND} algorithm is applied to each FOF group to identify locally overdense and self–bound substructures. The most massive of these is referred to as the "main", or smooth, subhalo, and is typically associated with the central galaxy, while smaller subhalos correspond to orbiting satellites or infalling systems.

For each particle type -- dark matter, gas, stars, and black holes -- \texttt{SUBFIND} determines membership based on local density peaks and gravitational binding energy. The resulting catalogs provide the physical and structural properties (e.g., total mass, stellar mass, gas fraction, and potential minimum) for both the main halo and its substructures. In this work, we base our halo shape analysis on all particles within the virialized region of each FOF group, rather than limiting the study to the main \texttt{SUBFIND} halo, using the minimum of the gravitational potential as the halo center. This choice ensures that the inferred three–dimensional shapes account for the full mass distribution, including substructures and diffuse components, which is essential when comparing different dark matter models such as CDM, SIDM, and WDM, where subhalo abundance and survival can vary significantly \citep[e.g.][]{giocoli08b,giocoli10a,despali22, ragagnin22,giocoli24,ragagnin24}. For this work, we make use of the particle positions and halo catalogs from the $z=0$ snapshot.

\begin{figure*}
\centering
\includegraphics[width=\hsize]{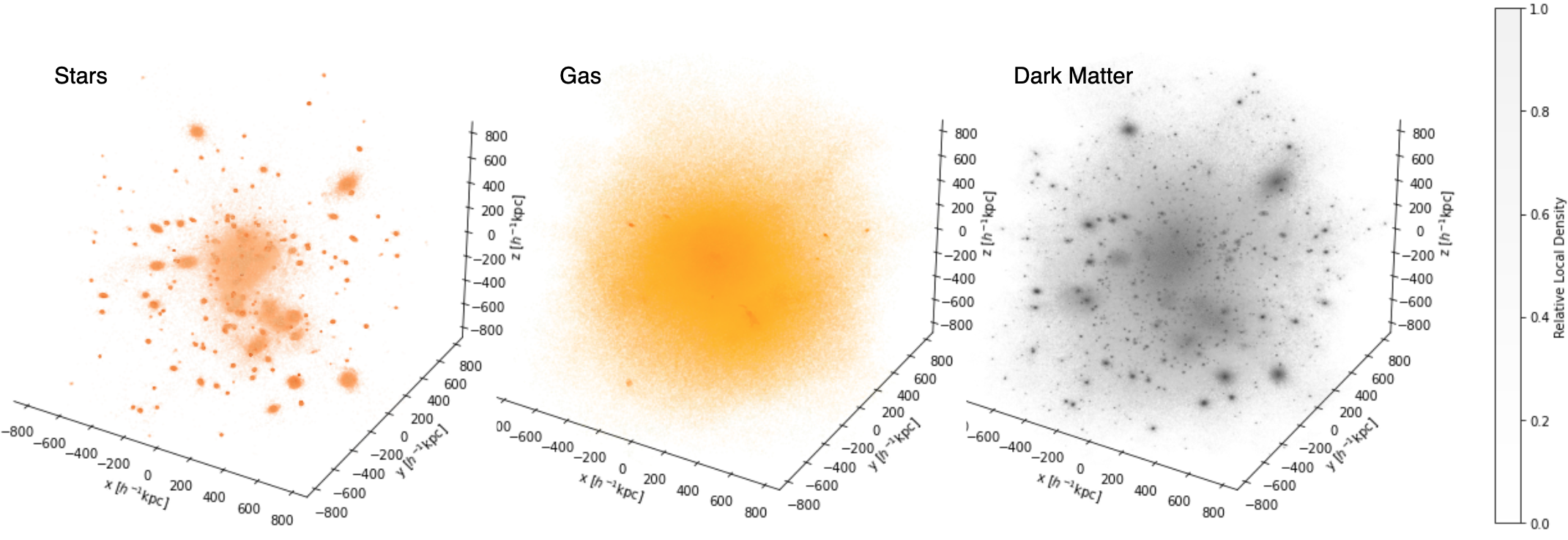}
\caption{Three-dimensional distribution of particles in stars (left), gas (center), and dark matter (right) of the most massive halo in the CDM simulation at $z=0$ having a mass  $M_{200}=2.6 \times 10^{14}h^{-1}$M$_{\odot}$.}
\label{fig_image}
\end{figure*}

\section{Global halo shapes}
\label{sec_globalshapes}
Numerous studies have demonstrated that the density distribution of dark matter halos is better described by triaxial ellipsoids rather than by assuming spherical symmetry \citep[e.g.][]{allgood06,peter13,despali13,despali17,chua19}. This triaxial nature reflects the anisotropic collapse and hierarchical assembly of structures in the cosmic web, where halos accrete matter preferentially along filaments and experience repeated mergers that preserve memory of large-scale anisotropies. Departures from spherical symmetry are therefore a natural outcome of structure formation in a $\Lambda$CDM Universe \citep{sheth01b,sheth02,musso21} and can be further modified by baryonic processes and the small-scale properties of dark matter \citep{blumenthal86,borgani06,cusworth14,cui16}.

In Fig. \ref{fig_image} we show the particle distribution in the most massive halo identified at $z=0$ in the CDM hydro simulation. The left, central, and right panels show the distribution of stars, gas, and dark matter, respectively. While the stars and dark matter appear clumpy, distributed between the main halo and dense substructure components, the gas is considerably smoother. It is therefore clear that the triaxiality of these components differs significantly, reflecting their distinct dynamical and dissipative properties.

\subsection{Shape parameters}
We use the inertia tensor to determine the shapes of all halos in the simulation, as well as of their baryonic components: gas and star particles. 
Commonly, for a system of $N$ particles with masses $m_i$ and position vectors $\mathbf{x}_i$ measured with respect to the halo center, the mass-weighted tensor is defined as
\begin{equation}
M_{\alpha \beta} = \dfrac{1}{N}  \sum_{i=1}^{N} \dfrac{m_i \, x_{i,\alpha} \, x_{i,\beta}}{r^2_i} \dfrac{1}{\sum_{i=1}^N m_i}\,,
\end{equation}
where $r_i$ denotes the ellipsoidal distance of each particle from the halo center:
\begin{equation}
r_i^2 = \dfrac{ \mathrm{d}  x_i^2}{a^2}  + \dfrac{  \mathrm{d} y_i^2}{b^2} +  \dfrac{\mathrm{d} z_i^2}{c^2}\,.
\end{equation}

The weighting in the inertial tensor ensures convergence toward a self-consistent ellipsoidal shape \citep{jing02,allgood06,zemp11}. 
Shape determination is performed iteratively: starting from a spherical volume, the tensor is computed, diagonalized, and the resulting axis ratios are used to redefine the ellipsoidal boundary, enclosing $N$ particles --- also updated at each step --  until a $1\%$ convergence on the axis ratios is achieved.
The resulting axis ratios describe the flattening of the halo \citep{bardeen86,matarrese91}.  If $a\ge b\ge c$, we define the ellipticity,
\begin{equation}
\epsilon = \dfrac{1-\left(\dfrac{c}{a}\right)^2}{2 \left[ 1 + \left(\dfrac{b}{a}\right)^2 + \left(\dfrac{c}{a}\right)^2 \right]}\,,
\end{equation}
and the prolateness,
\begin{equation}
p = \dfrac{1-2 \left(\dfrac{b}{a}\right)^2 + \left(\dfrac{c}{a}\right)^2}{2 \left[ 1 + \left(\dfrac{b}{a}\right)^2 + \left(\dfrac{c}{a}\right)^2 \right]}\,.
\end{equation}

We have prolate those halos where $p>0$; oblate halos where $p<0$. In the limiting cases where $p = \epsilon$ or $p = -\epsilon$, one obtains either a prolate or an oblate spheroid, respectively.  We can also quantify the degree of triaxiality using the parameter
\begin{equation}
T = \frac{1 - \dfrac{b^2}{a^2}}{1 - \dfrac{c^2}{a^2}},
\end{equation}
where $T = 0$ corresponds to a perfectly oblate spheroid, $T = 1$ to a prolate spheroid, and intermediate values indicate triaxial configurations \citep{franx91,bett07,despali13}. Typically, halos with $T < 0.33$ are classified as oblate, those with $T > 0.67$ as prolate, and the rest as triaxial.
These quantities provide self-consistent and reproducible measures of the intrinsic halo geometry, allowing for direct comparison across simulations with different dark matter models and baryonic physics implementations. In the context of the AIDA-TNG simulations, we apply this method to quantify the three-dimensional shapes of halos formed under cold, warm, and self-interacting dark matter scenarios, and to investigate how baryonic feedback and DM microphysics jointly affect their morphological evolution.

\begin{figure*}
\centering
\includegraphics[width=0.49\hsize]{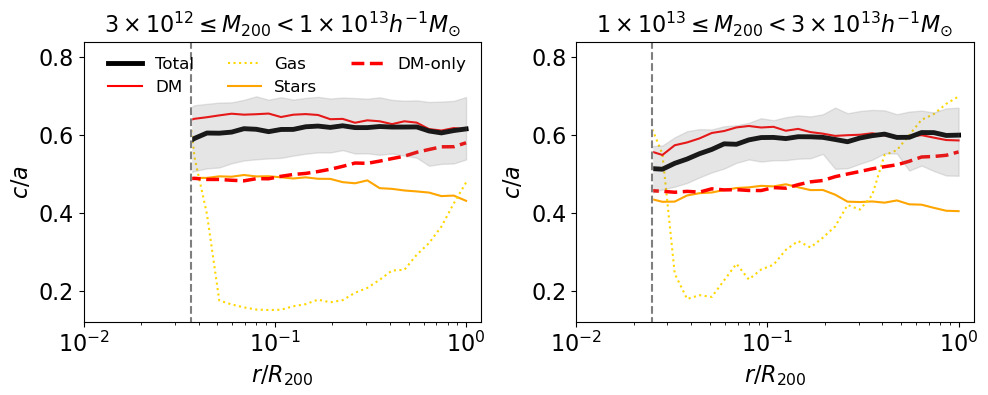}
\includegraphics[width=0.49\hsize]{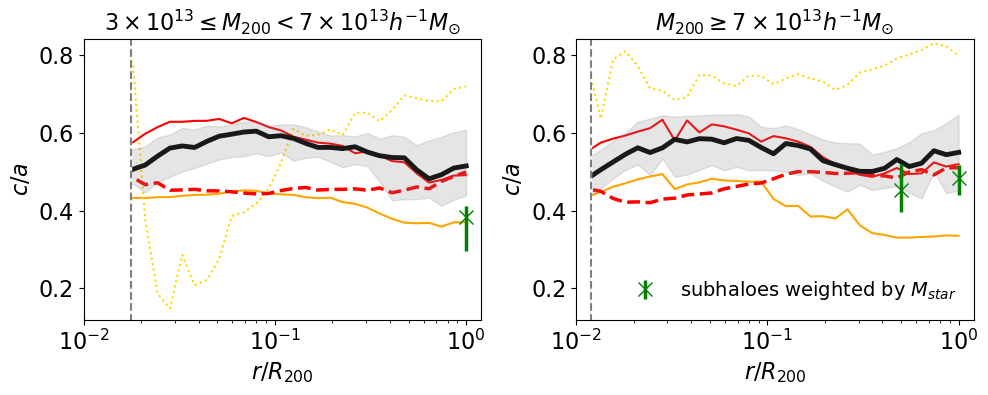}
\includegraphics[width=0.49\hsize]{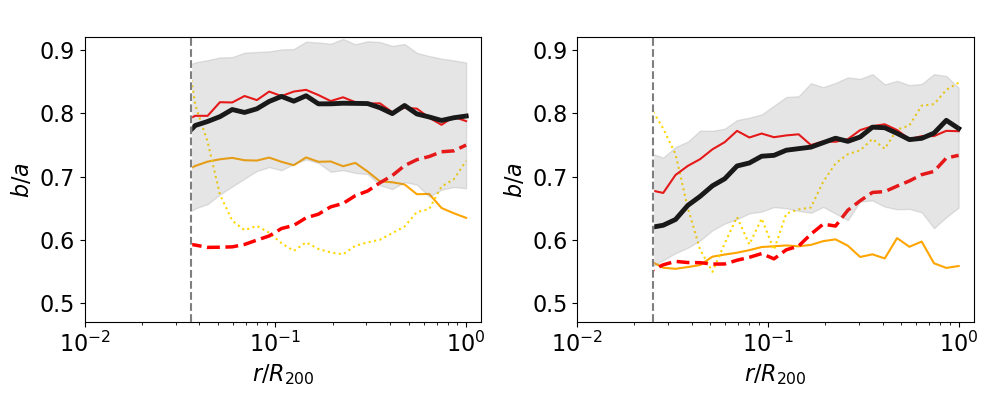}
\includegraphics[width=0.49\hsize]{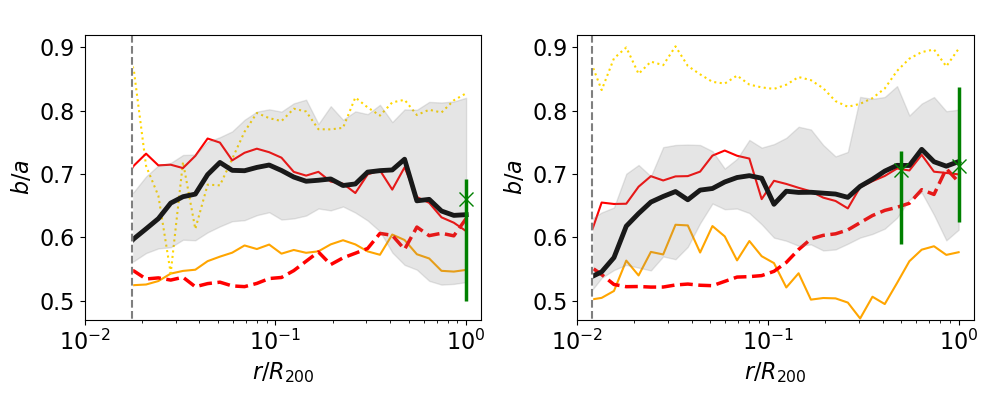}
\caption{Median profiles of minor-to-major (top) and intermediate-to-major (bottom) axis ratios for $z=0$ halos in the CDM simulation in bins of increasing mass (from left to right).
Solid black curves show the total axis ratios in the FP run, while solid red, dotted gold, and solid orange curves show the corresponding results for the dark matter, gas, and stars, respectively. Dashed red curves show halos in the same mass bins but for the DMO run. Green data points (shown only in some of the boxes) consider
the satellites and were weighted by stellar mass when computing the inertial mass tensor. The error bars bracket 25\% and 75\% of the distribution. The vertical bars display the resolution limit, which is equal to 10 times the softening length. The number of halos present in the four considered mass bins is: 305, 78, 34, and 14, respectively.}
\label{fig_LCDM_Radius}
\end{figure*}

We compute halo shapes, reading particle positions and accumulating them within 32 concentric radial bins, spaced logarithmically in the interval $-2 \leq \log(r/R_{200}) \leq 0$, where $R_{200}$ represents the radius enclosing 200 times the critical density of the universe. While \citet{despali22} restricted their shape measurements to the main smooth halo identified by \texttt{SUBFIND}, we include all particles bound within the virialized region, thereby capturing the full contribution from both the main halo and its substructures.
Shape measurements are performed independently at each enclosed radius, allowing the ellipsoid’s axis ratios and orientation to vary freely. To ensure robust shape estimates and define the smallest radius at which we trust the measurements of the various components, a minimum of 64 dark-matter particles per shell is required above 10 times the gravitational softening length of the simulation, following the convergence test done by \citet{chua19} -- as for the TNG-2 run that has the same resolution as the AIDA-TNG simulations.

We also estimate the shape from the satellite distribution, using the stellar mass as a weight (we require satellites to have at least 32 dark matter and 10 star particles), without iteration in this case. We only do this on two scales:  at $R_{200}$ and $R_{200}/2$, and we only consider enclosed radial bins that have at least 8 satellites when estimating the 3D shape. 

\subsection{Shapes in the CDM model}
In Fig.~\ref{fig_LCDM_Radius}, we present the median radial profiles of the intermediate-to-major ($b/a$, top panels) and minor-to-major ($c/a$, bottom panels) axis ratios for all mass components of halos at $z=0$ in the CDM scenario. We consider four distinct mass bins, as indicated at the top of each panel, containing 305, 78, 34, and 14 systems ordered in increasing mass. These profiles illustrate how the three-dimensional morphology of halos varies from the virial radius down to the innermost regions, and how this behavior depends on the baryonic content and feedback processes. In the DMO run, displayed using a red dashed style line, we recover the expected dependence of the axis ratios on distance, with a higher triaxiality towards the halo center \citep[e.g.][]{allgood06,despali17,chua19}. We underline that we use the solid red line style for the DM component of the FP run.

Across all mass ranges, the total matter distribution closely follows that of the dark matter from the virial radius down to approximately half this value, indicating that at these large scales the baryonic component follows the dominant dark matter potential. However, a mild deviation emerges at smaller radii: the dark matter becomes slightly more spherical than the total mass distribution. This difference reflects the influence of the dissipative baryonic component, which tends to settle into more compact structures within the central potential well, dominating the total density profile. 

Star particles display more triaxial shapes than the dark matter at nearly all radii. This enhanced triaxiality likely originates from the complex assembly history of the stellar component, which includes contributions from in-situ star formation as well as the accretion of satellite galaxies. The alignment of the stellar component with the underlying dark matter potential is therefore only partial, consistent with previous findings from hydrodynamical simulations \citep[e.g.][]{tenneti14,chua19}. In addition is worth mentioning that baryons make halos rounder and more oblate, and strengthen the alignment between the stellar and dark matter components \citep{chua22}. These effects become more pronounced with increasing stellar-to-halo mass ratio, indicating that systems with higher baryon conversion efficiency experience a stronger modification of their inner potential. This trend agrees with our AIDA–TNG results, where the inner halo becomes progressively more spherical as the baryonic mass fractions increase, highlighting the central role of feedback-regulated galaxy formation in shaping the dark matter distribution.
We discuss this in more detail in the next section.

It is worth noticing that the morphology of the gaseous component within dark matter halos reflects a complex interplay between gravitational dynamics and baryonic feedback. 
As shown in the top panels, the minor-to-major axis ratio $c/a$ indicates that gas tends to concentrate toward the center in low-mass halos, resulting in a more oblate configuration. 
In contrast, in more massive systems, this central condensation is almost completely suppressed, most likely due to the impact of strong AGN feedback that counteracts cooling and prevents gas from accumulating efficiently in the central regions. This transition highlights the competing roles of radiative cooling and feedback in shaping the internal structure of halos across the mass spectrum. It is worth noting that, toward the center, the gas particles tend to follow the stellar shape, concentrated in the central galaxy.

Interestingly, the shapes derived from the spatial distribution of satellite galaxies -- green data points with error bars -- provide a good proxy for the global halo morphology. This result is particularly relevant for observations, as the spatial anisotropy of satellite systems can be directly measured in galaxy surveys \citep[e.g.][]{brainerd05,tempel15}. Therefore, the agreement between the satellite-based and total matter shapes suggests that observational measurements of satellite distributions can be used to constrain the intrinsic triaxiality of dark matter halos in the real Universe.

Overall, Fig.~\ref{fig_LCDM_Radius} demonstrates that halo shapes exhibit systematic and physically motivated variations with radius, component, and halo mass. While the anisotropic accretion of dark matter largely governs the outer regions, the central morphology is strongly modulated by baryonic physics — specifically, cooling, star formation, and feedback — which collectively determine how matter settles into equilibrium within the central potential well.

\subsection{The effect of SIDM and WDM on shapes}
As discussed by \citet{despali22} and \citet{mastromarino23}, alternative dark matter models introduce additional modifications to the picture described above. Self-interactions are expected to lead to isotropic cores enhancing sphericity within the central regions, while the smoother accretion histories of warm dark matter halos may result in rounder outer profiles. These distinct morphological imprints provide a promising avenue for discriminating between dark matter scenarios, particularly when combined with observables sensitive to halo geometry and alignment. 

In the context of the AIDA-TNG simulations, the consistency of these trends across a wide range of masses and redshifts suggests that halo shape statistics can serve as a robust diagnostic for both baryonic feedback strength and the nature of dark matter. 

\begin{figure*}
\centering
\includegraphics[width=0.49\hsize]{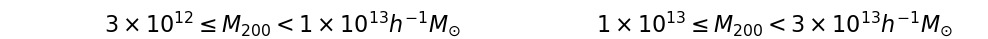}
\includegraphics[width=0.49\hsize]{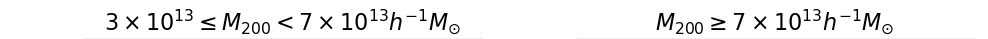}
\includegraphics[width=0.49\hsize]{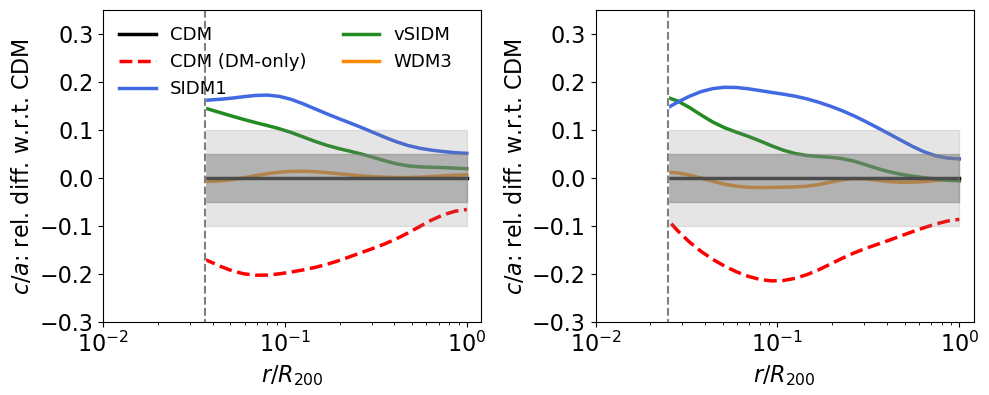}
\includegraphics[width=0.49\hsize]{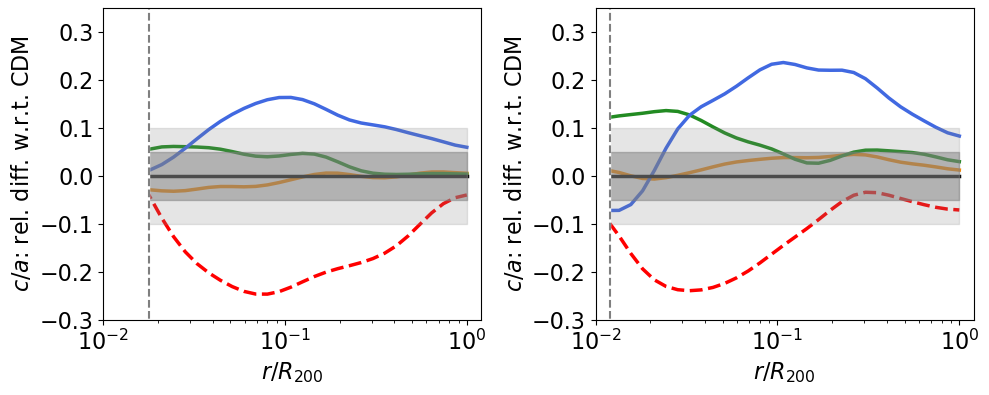}
\includegraphics[width=0.49\hsize]{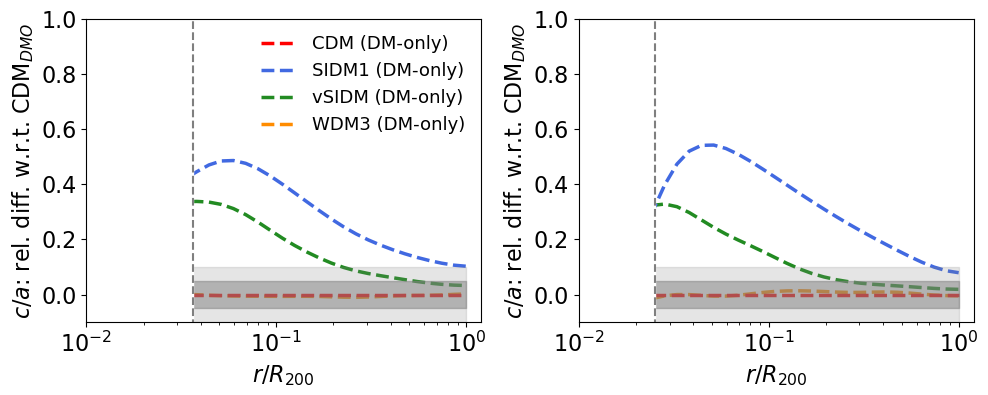}
\includegraphics[width=0.49\hsize]{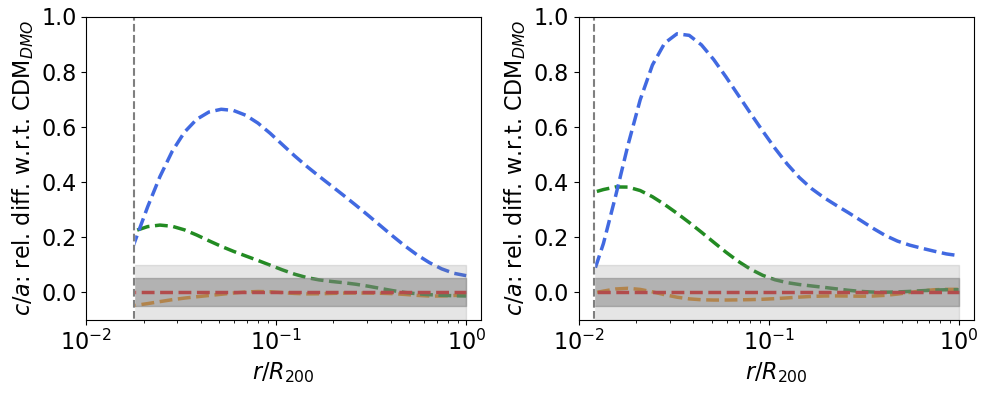}
\caption{Minor-to-major axis ratios for the DM models relative to the CDM value, for the same ($z=0$) mass bins as Fig. \ref{fig_LCDM_Radius}. Top and bottom panels display the results from FP and DMO runs, respectively.}
\label{reldiff_c_over_a}
\end{figure*}

In the top panels of Fig. \ref{reldiff_c_over_a}, we compare the total minor-to-major axis ratio of the two self-interacting and warm dark matter models with respect to the corresponding CDM reference hydro-run. The dark and light gray shaded regions in all panels indicate 5\% and 10\% differences, respectively. The different panels refer to the same mass bins presented in Fig. \ref{fig_LCDM_Radius}.
This comparison offers valuable insights into how both microphysical and astrophysical processes contribute to the overall morphology of halos across different radial scales. In dashed red we display the case of the CDM DMO simulation: as discussed in the previous section, halos are generally more triaxial and elongated in the inner regions, consistent with previous findings from large-volume cosmological runs \citep[e.g.][]{allgood06,despali17,chua19}, and small-scale Milky Way-size halos \citep{chua21}. 

In the bottom panels of Fig. \ref{reldiff_c_over_a}, we display the relative differences of $c/a$ but for the DMO runs. This can serve as a guide to better understand and quantify the effects of baryonic physics processes that additionally modify the radial profile of the three-dimensional shape of halos of different masses. The changes introduced by ADM and baryons are qualitatively distinct.

Elastic scattering between dark matter particles transfers momentum and heat within the halo, leading to the formation of a central core and a substantial redistribution of momentum through particle scattering \citep[e.g.][]{rocha13,spergel00,robertson21}. Consequently, SIDM halos are expected to be more spherical than their CDM counterparts, particularly within the inner $\sim 0.1\, R_{\mathrm{200}}$. 

SIDM halos, both for constant and scale-dependent cross sections, are more spherical than CDM in the DMO run (bottom panels of Fig. \ref{reldiff_c_over_a}). This trend, albeit diminished, persists when baryons are included, although the combined effect depends on the efficiency of feedback and the characteristic cross-section of the interaction. In the top panels, we see that the baryonic and feedback mechanisms influence and reduce the differences between CDM, SIDM, and vSIDM, especially in the central regions. In all panels, the WDM3 is relatively close, within 5\%, to the CDM case. The small differences are due to the small-scale suppression, late forming, and less concentrated WDM halos. For the two largest mass bins -- the last two column subpanels -- we can notice that at 10\% the halo boundary radius, there is a clear difference between SIDM1 and vSIDM models, where the halos in the former are more spherical than in the latter. However, at smaller radii, the behavior changes.  While vSIDM halos remain more spherical than their CDM counterparts, SIDM1 halos exhibit shapes that are comparable to those in CDM. These trends are qualitatively similar in both the DMO and FP runs, underlining that in the latter, the shape toward the center is more strongly influenced by baryonic physics mechanisms \citep{despali22}. 

These systematic differences in halo morphology across models have direct observational implications. The three-dimensional shape of dark matter halos affects gravitational lensing observables, such as the projected ellipticity of convergence maps and the azimuthal variation of weak-lensing shear profiles \citep[e.g.][]{meneghetti01,meneghetti07a,oguri10,bonamigo15,clampitt16,despali17}. Furthermore, the internal alignment between baryonic and dark matter components influences galaxy–halo misalignment statistics \citep{piras18, xu23a}, which can be probed via intrinsic alignment measurements in weak-lensing surveys \citep{tenneti15,hildebrandt17,heymans21,deskids23}. By linking the morphological differences observed in AIDA-TNG to these measurable quantities, we aim to provide a physically motivated framework for interpreting future constraints from high-precision lensing and clustering analyses, and to test the viability of alternative dark matter models in the era of Euclid and Rubin.

\section{Misalignment angles between star, gas, and dark matter components}
\label{sec_misal}
The spatial and dynamical alignment between the stellar, dark matter, and gas components provides an essential diagnostic of halos' internal structure and assembly history. In $\Lambda$CDM cosmologies, the stellar and dark matter distributions are found to have similar orientations within the central regions, with typical misalignment angles of $\sim 15$ -- $30^{\circ}$ depending on halo mass, morphology, and feedback strength \citep[e.g.][]{bett10, velliscig15, chua19, despali22}. Recent work suggests that the accreted stellar mass fraction is a key factor in determining the misalignment \citep{xu23b}. The gas component, being more dynamically responsive to feedback and accretion processes, often shows larger offsets, particularly in systems with strong AGN activity or recent mergers.

\begin{figure*}
\centering
\includegraphics[width=0.49\hsize]{figures/top_cut-0.jpg}
\includegraphics[width=0.49\hsize]{figures/top_cut-1.jpg}
\includegraphics[width=0.49\hsize]{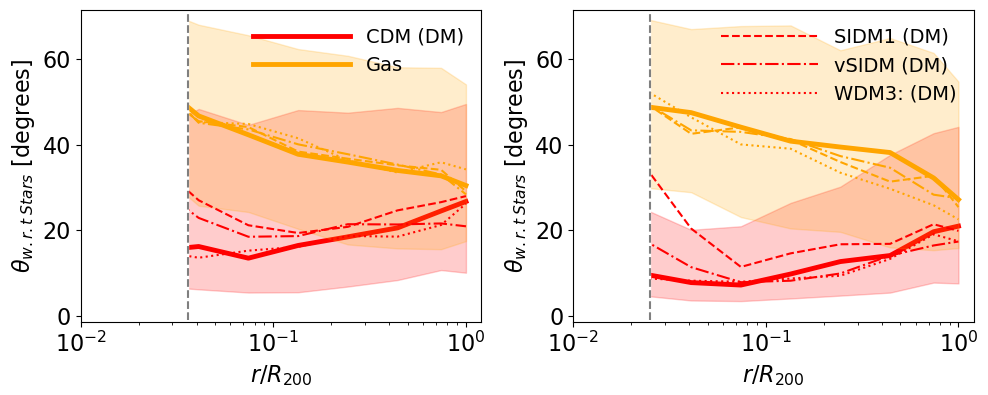}
\includegraphics[width=0.49\hsize]{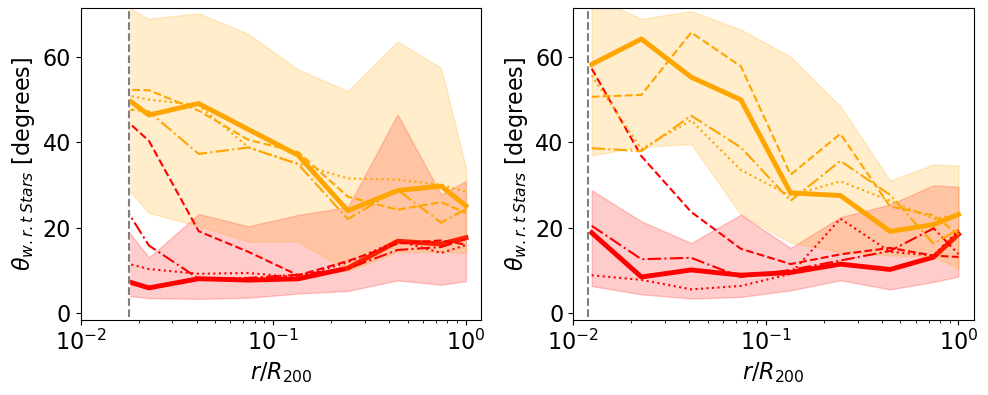}
\caption{Median misalignment angles, as a function of the rescaled radius, of dark matter (red) and gas (orange) with respect to the star particles (yellow), in the AIDA-TNG simulations.  Dashed, dot-dashed, and dotted lines refer to SIDM1, vSIDM, and WDM3 simulations, respectively, and the four panels are for the same mass bins as in Fig. \ref{fig_LCDM_Radius}. The colored shaded regions bracket the central 50\% (i.e., between 25\% and 75\%) of the CDM measurements. The four panels refer to various mass bins as presented in Fig. \ref{fig_LCDM_Radius}
\label{fig_misalignment}}
\end{figure*}

In the AIDA-TNG simulations, we compute the misalignment between components by measuring the angle between the major axes of the corresponding inertia tensors, derived for the stellar with respect to the dark matter and gas particle distributions. This is done in light of considering optically selected stellar components from photometric observations.  
This approach allows us to quantify the three-dimensional orientation of each component as a function of radius and halo mass. Results are displayed in Fig. \ref{fig_misalignment}. Red and orange lines represent the misalignment of dark matter and gas components with respect to the stars. CDM, SIDM1, vSIDM, and WDM3 are represented by solid, dashed, dot-dashed, and dotted lines, respectively. The four panels refer to various mass bins as presented in Fig. \ref{fig_LCDM_Radius}. 

A systematic difference emerges when comparing self-interacting with cold dark matter models. SIDM halos tend to be more spherical in their inner regions, as frequent dark matter particle scatterings isotropize the velocity field and redistribute mass \citep[e.g.][]{rocha13, elbert15, robertson21}. This enhanced isotropy reduces the coupling between the stellar and dark matter shapes, resulting in larger misalignment angles in the central regions compared to CDM, particularly in the SIDM1 model. For the same reason, the orientation of the stellar component -- largely shaped by baryonic inflows and feedback -- is less constrained by the underlying halo. It is worth underlining that the effect is stronger in more massive halos, as shown in the right panels of the figure. Large misalignment angles can also be due to the fact that when shapes are rounder the direction of the axes are less constrained. This is the case for the SIDM1 dark matter shapes at the high-mass end, as well as for the gas distribution in all mass bins.

\begin{figure}
    \centering
    \includegraphics[width=\linewidth]{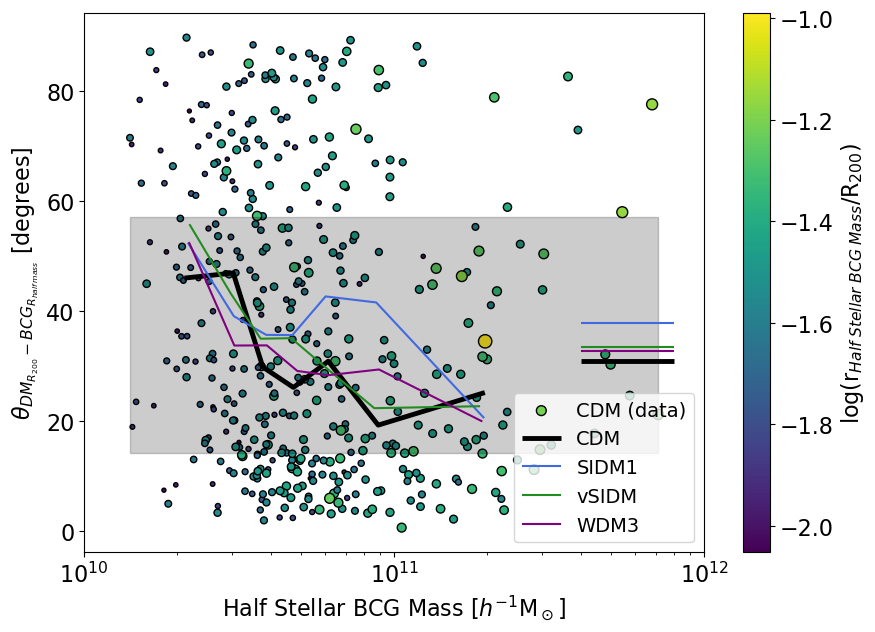}
    \caption{Misalignment angle between the major axis of the BCG (estimated using the inner 50\% of the star particles) and of its host dark matter halo (estimated from the total mass within $R_{200}$), shown as a function of half stellar mass of the BCG. Symbol size and color are proportional to the ratio between the central galaxy and host halo sizes for the CDM runs.
    A dashed gray band encloses 50\% of the data points.
    Black, blue, green, and purple lines show the running median for the CDM, SIDM1, vSIDM, and WDM3 cases, respectively. Horizontal lines on the right indicate the median value, over all half stellar mass values, for the corresponding DM model.}
    \label{theta_BCG}
\end{figure}

The gas component displays a more complex behavior. In massive systems, where AGN feedback dominates, the gas can become uncorrelated with both the stellar and dark matter components. In contrast, in lower-mass halos, the gas orientation is less misaligned with respect to the overall potential. This results in a weak dependence of the gas–dark matter misalignment on halo mass but a dependence on the underlying dark matter model.

Overall, the misalignment statistics provide a complementary probe of the nature of dark matter and its correlation with the formation processes and assembly histories of halos \citep{donahue16}. While CDM predicts a tight alignment between stars and dark matter, SIDM models naturally produce larger orientation scatter and rounder central morphologies. Future comparisons with observational constraints from weak and strong lensing studies \citep[e.g.][]{schaller15b,schaller15c, harvey19, harvey21a,harvey21b, robertson23} and from integral field spectroscopy of stellar kinematics may therefore help distinguish between these dark matter scenarios.

In this respect, in Fig. \ref{theta_BCG}, we show the misalignment angle between the bright central galaxy and the host dark matter halo. We measure this angle between the major axis of the mass tensor ellipsoid, as measured at half-stellar mass and at the halo boundary radius, quantifying how well the central galaxy traces the shape of the host dark matter halo.
The misalignment between the BCG (bright central galaxy, even when referring to a more general central galaxy) and its host dark matter halo provides a direct probe of the coupling between baryons and the underlying gravitational potential. In hierarchical structure formation, the major axes of halos and their central galaxies are expected to align along the dominant filamentary accretion direction in the formation phase (e.g. \citealt{tenneti15, chisari17, okabe20a}). However, baryonic feedback processes, mergers, and dark matter self-interactions can perturb this alignment. 
From the figure, we notice that CDM halos tend to be slightly more aligned than the SIDM ones, rounder inner halos, and larger angular offsets due to isotropization of the dark matter potential (e.g. \citealt{harvey19, robertson21}). The horizontal lines in the right part of the figure mark the median values, over the whole halo sample, of the misalignment angle for the different DM models; precisely equal to 31$^{\circ}$, 38$^{\circ}$, 34$^{\circ}$, and 33$^{\circ}$ for CDM, SIDM1, vSIDM and WDM3, respectively. The shaded gray region brackets 50\% of the CDM data points.
The misalignment is inversely related to the half-stellar mass, the host halo mass $M_{200}$, and to the ratio between the half stellar mass radius and $R_{200}$, highlighting that halos that form later have a smaller misalignment between the BCG and host dark matter halo. 
From an observational perspective, the distribution of BCG–halo misalignment angles is crucial for interpreting stacked weak-lensing ellipticities and intrinsic alignment statistics in upcoming wide-field weak lensing surveys. 

\section{Mass dependence of the minor to major axis ratio}
\label{sec_massdep}
The shape of dark matter halos encodes valuable information on their assembly history and the underlying physical processes that shape their evolution. In this section, we present the minor-to-major axis ratio ($c/a$) of halos as a function of their virial mass, comparing results from the CDM DMO simulation and the FP AIDA-TNG runs.

\begin{figure*}
\centering
\includegraphics[width=0.32\hsize]{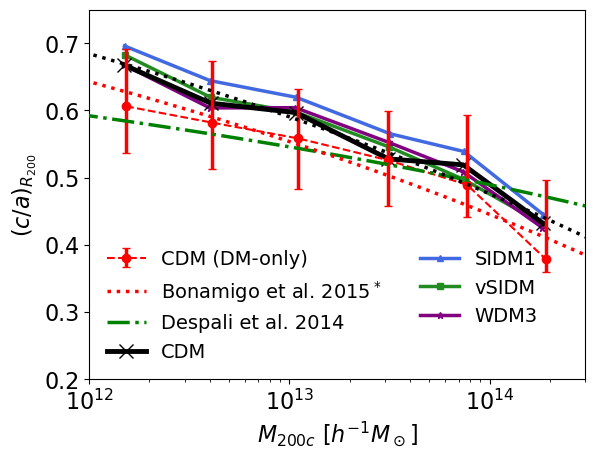}
\includegraphics[width=0.32\hsize]{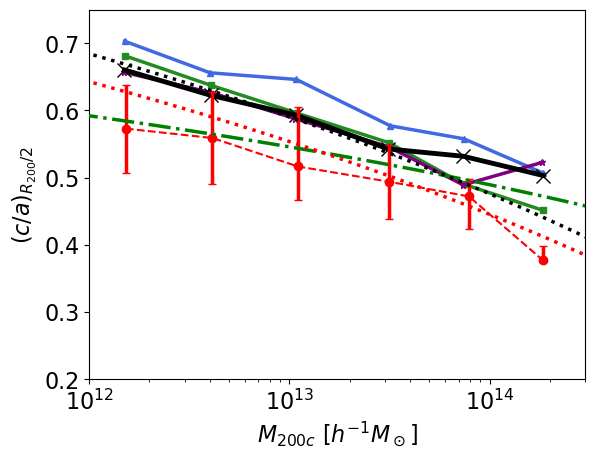}
\includegraphics[width=0.32\hsize]{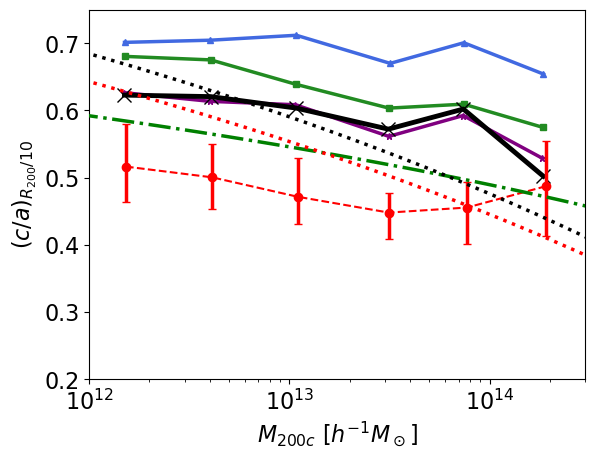}
\caption{\label{fig_c_over_c_M200}Minor-to-major axis ratios as a function of the halo mass $M_{200}$. Red data points display the median relation for the CDM DMO simulation at $z=0$, while the corresponding 
solid (dotted) error bars bracket $25$ and $75\%$ of the distribution at fixed halo mass. Black, blue, green, and magenta connected data points show the results for the FP runs in different dark matter models. To avoid overcrowding, these are displayed without the error bars.  Left, central, and right panels show the measurements at the halo boundary radius $R_{200}$, $R_{200}$/2, and $R_{200}$/10, respectively. The green dot-dashed curve shows the results from \citet{despali14}, while the red and black dotted lines show the best fit relation assuming the median value of the \citet{bonamigo15} model at $\left(c/a\right)_{R200}$ -- repeated in all three panels to guide the reader.}
\end{figure*}

\begin{figure*}
\centering
\includegraphics[width=0.32\hsize]{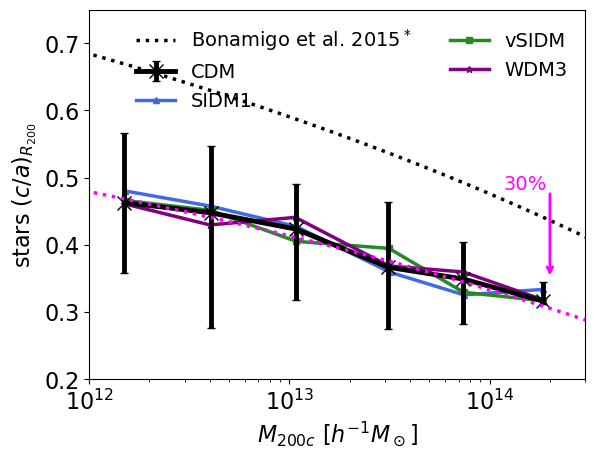}
\includegraphics[width=0.32\hsize]{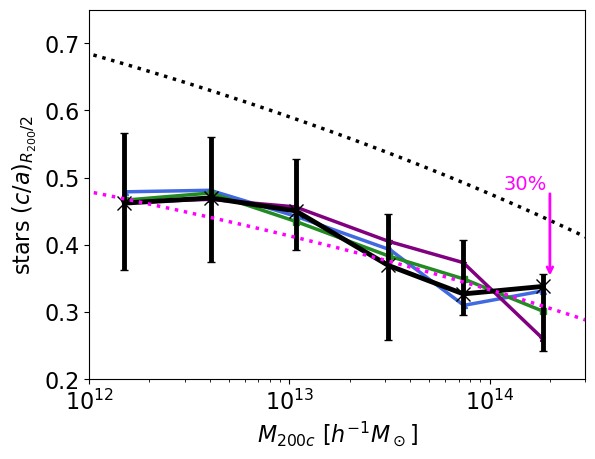}
\includegraphics[width=0.32\hsize]{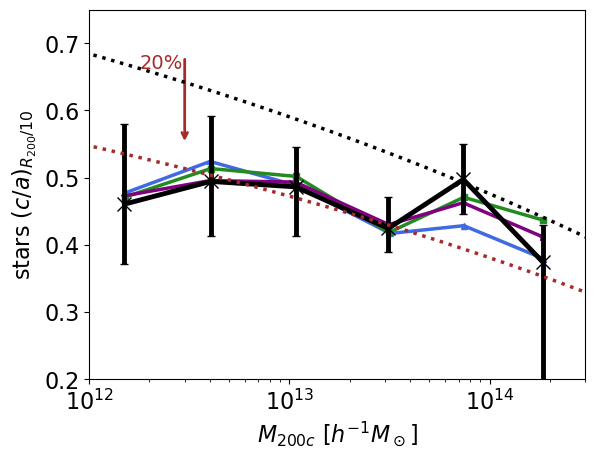}
\caption{\label{fig_c_over_c_M200_STARS}
Median minor-to-major axis ratios as a function of the halo mass $M_{200}$, considering only star particles for halos at $z=0$. Black, blue, green, and magenta connected data points exhibit the results for the FP runs considering different dark matter models. Error bars bracket $25$ and $75\%$ of the distribution, displayed only for the CDM case. Left, central, and right panels show the measurements at the halo boundary radius $R_{200}$, at its half, and $10\%$ of the value, respectively. The black dotted curve shows the recalibrated \citet{bonamigo15} model for the $\left(c/a\right)_{R200}$ as measured from all particles in the CDM run. The arrows in each panel indicate the percentage shift of the total-matter $c/a$ relation required to match the minor-to-major axis ratios of the stellar component as a function of halo mass.}
\end{figure*}

In the DMO case, the median $c/a$ ratio shows a clear mass dependence, with more massive halos being systematically less spherical. This trend reflects the stronger influence of anisotropic mass accretion and recent mergers in high-mass systems \citep{allgood06, despali14, bonamigo15}. Low-mass halos, which typically form earlier and have smoother accretions, have had more time to relax dynamically and therefore display rounder configurations. These results are shown in Fig. \ref{fig_c_over_c_M200}, where the red points and error bars show the median of the measurements in the CDM DMO run at $z=0$, solid (dotted) error bars bracket $25$ and $75\%$ of the distribution at fixed host halo mass. Left, central, and right panels display the minor-to-major axis ratios as measured at the halo boundary radius $R_{200}$, at $R_{200}/2$, and at $R_{200}/10$, respectively. Our measurements in the CDM DMO run are in good agreement with the relations derived by \citet{despali14} (dot-dashed green curve, identical in all panels of the figure) and \citet{bonamigo15}, confirming that the AIDA-TNG suite reproduces the well-established shape–mass dependence found in previous cosmological simulations.

Following \citet{bonamigo15}, the median relation  can be computed from the probability density function for the Beta:
\begin{equation}
f\left(\dfrac{c}{a}, \alpha, \beta  \right) = \dfrac{\Gamma(\alpha+\beta)}{\Gamma(\alpha) \Gamma(\beta)}\,
\left(\dfrac{c}{a}\right)^{\alpha-1}
\left(1 - \dfrac{c}{a}\right)^{\beta-1}\,,
\end{equation}
where $\Gamma(x)$ represents the gamma function\footnote{We have computed this using \texttt{scipy.stats.beta}.}.
\citet{bonamigo15}, using the \texttt{SBARBINE} \citep{despali16} and the \texttt{Millennium XXL} \citep{angulo12} simulations, have shown that the free parameters $\alpha$ and $\beta$ depend on the peak-height $\nu$ of the corresponding halo virial mass $M_{vir}$ as follows:
\begin{eqnarray}
\alpha(\nu) &=& \dfrac{\beta(\nu)}{\mu(\nu)^{-1}-1}\,,\\
\beta(\nu) &=& 0.560 \log(\nu) + 0.836 \,,\\
\mu(\nu) &=& m \log(\nu) + q\,,
\label{eq:mulognu}
\end{eqnarray}
where $m=-0.322$ and $q=0.620$. However, to improve the fit to our DMO and FP CDM simulations, we run an MCMC analysis and let those parameters vary within the ranges $[-2, -1]$ and $[0, 1]$, respectively. When converting $M_{vir}$ to $M_{200}$, we assume that halos are described by an NFW profile \citep{navarro96,navarro97} with mass and concentration related by the \citet{ludlow16} model\footnote{In this implementation, we use \texttt{COLOSSUS} \citep{diemer18}}. The red dotted curve -- same in all panels -- represents the best fitting \citet{bonamigo15} model to the DMO CDM simulation at $z=0$, with $c/a$ computed at the halo boundary radius $R_{200}$.  This fit has 
\begin{equation}
    (m,q) = \left(-0.522^{+0.210}_{-0.209}, 0.591^{+0.042}_{-0.043}\right)
\end{equation}
for the linear relation parameters in Eq.~(\ref{eq:mulognu}).

Black, blue, green, and magenta symbols (connected by lines), in Fig.~\ref{fig_c_over_c_M200} show the results of the CDM, SIDM1, vSIDM, and WDM3 hydro runs, respectively. 
When baryonic physics is included, the halo shapes become significantly rounder, particularly at low and intermediate masses, as well as in the central regions. The median $c/a$ ratio in the FP AIDA-TNG run increases by approximately $15$–-$20\%$ compared to the DMO case, depending on the mass range considered. This sphericalization is a well-known outcome of baryonic condensation in the central regions of halos \citep[e.g.][]{kazantzidis04, zemp11, zemp12, despali22}. The effect is strongest in halos with $M_{200} \lesssim 10^{13},h^{-1}M_\odot$, where stellar and gaseous components dominate the inner potential, while for more massive halos the efficiency of AGN feedback limits baryonic contraction, thereby preserving a more triaxial configuration \citep{bryan13, butsky16, chua19}. The black dotted curve is the best fit relation for the full-hydro CDM run at $z=0$ for $c/a$ measured at $R_{200}$, with:
\begin{equation}
    (m,q) = \left(-0.553^{+0.233}_{-0.235},0.629^{+0.051}_{-0.050}\right).
\end{equation}

The comparison between the DMO and FP AIDA-TNG runs suggests that baryonic processes primarily shift the normalization of the $c/a$–$M_{200}$ relation upward, rather than altering its slope. This indicates that while the cosmological assembly history drives the overall dependence of halo shape on mass, baryons impose an additional, approximately mass-independent sphericalization effect modulated by feedback strength. This trend is consistent with what has been reported in other large-volume hydrodynamical simulations, such as Illustris and TNG \citep{chua19, despali22}.

In addition to baryonic physics, the AIDA-TNG simulations provide an ideal test bed for exploring the impact of alternative dark matter models on halo shapes. In particular, the SIDM and WDM runs allow us to disentangle the structural effects of self-interactions and free-streaming suppression from those due to baryonic feedback. As can be seen from Fig. \ref{fig_c_over_c_M200}, the difference between the various DM models becomes more evident as we move closer to the halo center. In particular, the SIDM models move upward with respect to the CDM model; their halos tend to be more spherical. 

In Fig. \ref{fig_c_over_c_M200_STARS} we show the median minor-to-major axis ratio relation considering only star particles, at $R_{200}$, $R_{200}/2$ and $R_{200}/10$ (left, middle, right-hand panels, respectively), with different colors representing the different DM models. The dotted black curve shows the recalibrated \citet{bonamigo15} model for $\left( c/a \right)_{R200}$ measured from all particles in the CDM run at $z=0$. From the figure, we can notice that star particles at $R_{200}$ and $R_{200}/2$ are typically $30\%$ more triaxial than the whole matter, while at $10\%$ of the halo boundary radius -- close to the BCG -- the star distribution is rounder (axis ratios are closer to unity).  The arrows illustrate how much the $c/a$  relation of the total matter must be shifted to describe the stellar minor-to-major axis ratios as a function of halo mass.

\begin{figure*}
\centering
\includegraphics[width=0.32\hsize]{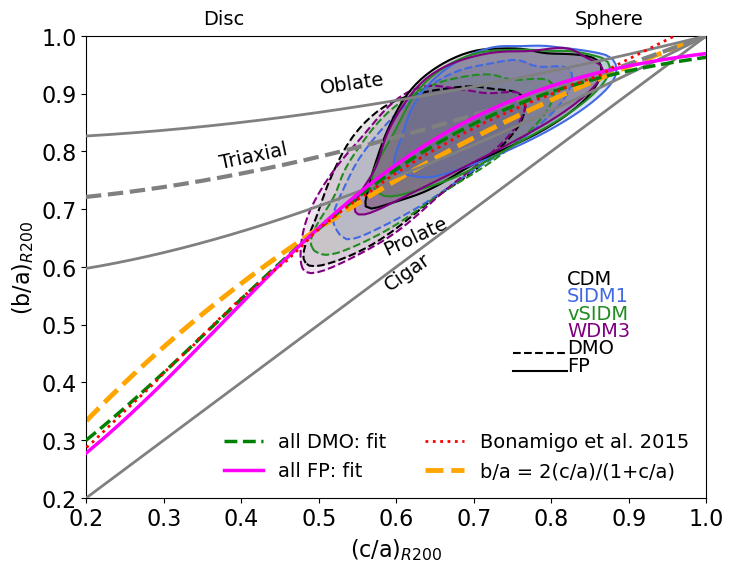}
\includegraphics[width=0.32\hsize]{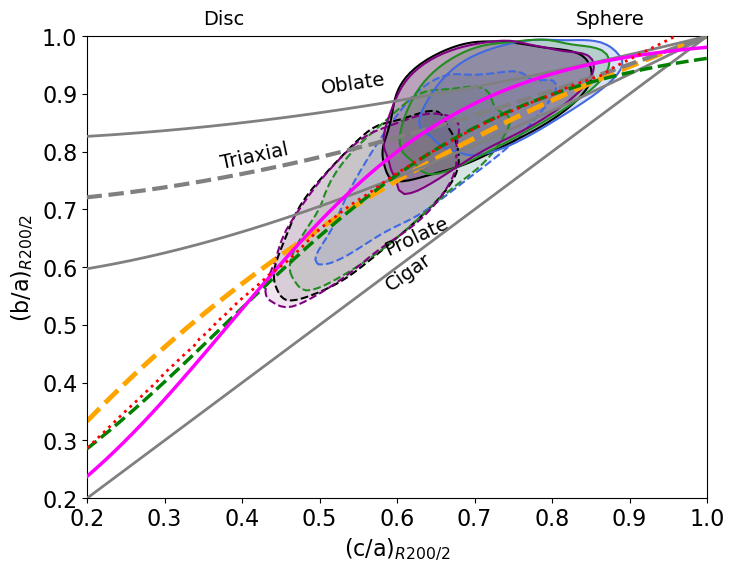}
\includegraphics[width=0.32\hsize]{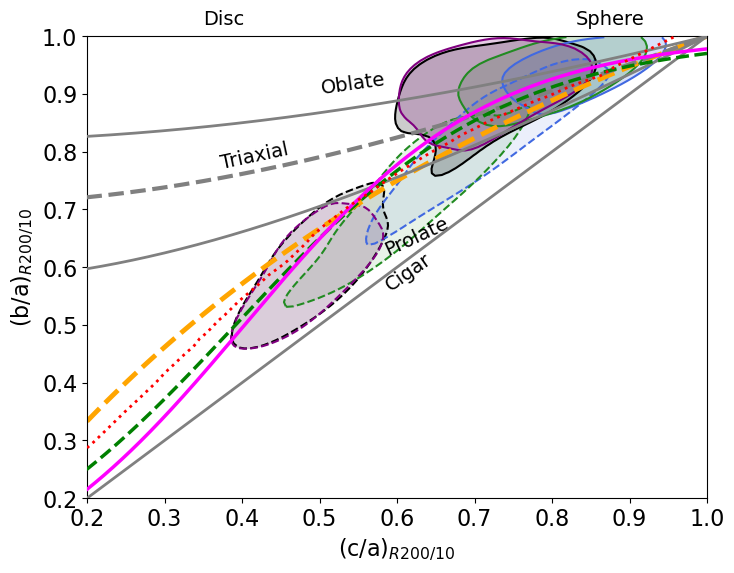}
\caption{\label{fig_c_over_a_b_over_a}Correlation between intermediate-to-major and minor-to-major axis ratio for halos at $z=0$ in the AIDA-TNG simulation.  Left, central, and right panels refer to the correlation measured at the halo boundary radius $R_{200}$, $R_{200}$/2, and $R_{200}$/10, respectively.
Shaded contours in the background represent the joint distribution of $b/a$ and $c/a$ (as measured considering all particles within $R_{200}$ and averaging over all hydrodynamic runs for the different DM models).  The blue line shows the median $b/a$ at fixed $c/a$, with error bars showing the range that encloses $25\%$ of the distribution on either side of this median. The black dotted line shows the corresponding median in the DMO runs.  The grey lines refer to the cases of halos with $T=1/3, 2/3$ and $1$ (top to bottom). The dashed gray line shows the case $p=0$, while the long-dashed gold curve shows the simplest prediction (for Gaussian initial conditions) described in the text, and the red dotted curve is the relation from \citet{bonamigo15}. 
}
\end{figure*}

\section{Correlation between intermediate and minor axis ratios}
\label{sec_correlation}
The joint distribution of the minor-to-major ($c/a$) and intermediate-to-major ($b/a$) axis ratios provides deeper insight into the intrinsic triaxiality of dark matter halos. While the minor-to-major axis ratio quantifies the overall flattening of a halo, the intermediate-to-major axis ratio captures its degree of prolateness or oblateness. Their correlation thus encodes the balance between anisotropic accretion, mergers, and internal relaxation processes that shape halo morphology \citep{allgood06, despali14, bonamigo15}.

In Fig.~\ref{fig_c_over_a_b_over_a}, we show the median distribution of $b/a$ versus $c/a$ for the halos identified in the hydrodynamical and dark matter-only AIDA-TNG runs, measured at $R_{200}$, $R_{200}$/2, and at $R_{200}$/10 from left to right, respectively.  Previous works \citep{jing02,rossi11} suggested that this relation should be quite universal. The different colored closed contours encircle 50\% of the data points in the various dark matter scenarios, while the solid and dashed line styles refer to the FP and DMO runs, respectively.   
From the figure, we can notice that while at $R_{200}$ the different dark matter models line within a percent or so, moving toward small radii, the SIDM models deviate with respect to the reference CDM case, while WDM3 remains close to the latter. 

The inclusion of baryonic physics significantly alters this relation. In the hydrodynamical AIDA-TNG runs, halos shift toward higher $c/a$ and $b/a$ values, indicating a transition from prolate to more oblate or nearly spherical configurations: the condensation of baryons makes halos rounder and more oblate \citep{chua22}. To get some intuition, the solid grey curves show the cases of halos with $T=1/3, 2/3$ and 1 (top to bottom), and the dashed grey curve shows $p=0$.  Clearly, halos with $c/a \ge 2/3$ lie on the $p=0$ relation, whereas those with $c/a\le 0.5$ tend to be prolate rather than triaxial.  

The simplest prediction for this relation, based on Gaussian random field statistics, has $b/a = 2 \left(c/a\right)/\left(1 + c/a\right)$
\citep{jing02,rossi11}.  
This tends to overpredict the value of $b/a$, especially when $c/a$ is small.  I.e., final halo shapes are more prolate than this simplest prediction, suggesting that anisotropic infall along cosmic filaments tends to elongate the halo along the preferred mass accretion direction \cite[e.g.][]{bett07, vera-ciro11, bonamigo15}.  

Following \cite{bonamigo15}, we found that a two-parameter logistic function provides a good fit:  
\begin{equation}
\dfrac{b}{a} = 1/\left\{ 1 + \exp{\left[-p_1 \left( \dfrac{c}{a} - p_0\right)\right]}\right\}\,.
\label{eqthirdorder}
\end{equation}
In order to compute the best fit parameters for $p_0$ and $p_1$, we bin all the corresponding dark matter models -- for the DMO and FP runs -- along $c/a$ value in 12 bins from 0.1 to 1, computing the median $b/a$ and two percentiles enclosing 25\% and 75\% of the distribution.

The dashed green and solid magenta curves show the best fit relation, for each of the three panels of Fig. \ref{fig_c_over_a_b_over_a}, to the median data points from all FP runs, the values for $p_0$ and $p_1$ are reported in Table \ref{tab_par} with the the best-fit reduced chi-squared $\chi^2_{\nu}$. For the values computed at the halo boundary $R_{200}$, the DMO relation is in excellent agreement with the relation reported by \cite{bonamigo15}, shown here by the dotted red curve.  

From the figure, it is also worth noting that the majority of halos are triaxial.
While the DMO systems are more prolate toward the center, clearly, in the FP AIDA-TNG runs, $b/a$ shifts to larger values when $c/a \geq 0.5$, indicating a transition from prolate to more oblate or axi-symmetric, if not simply more spherical configurations. This effect is strongest for intermediate-mass systems ($M_{200}\sim10^{12-13}h^{-1}M_\odot$), where baryonic condensation and feedback efficiently isotropize the potential \citep{kazantzidis04, butsky16, chua19, despali22}. 
Overall, the $b/a$–$c/a$ relation provides a compact and physically intuitive diagnostic of the mechanisms shaping halo morphology. The comparison across the AIDA-TNG dark matter models indicates that baryonic physics mainly contribute to halo rounding, albeit through different channels: baryons through dissipative condensation and feedback-induced relaxation, and in addition, SIDM through collisions tend to thermalize the dark matter component, which is very evident moving toward the halo center. These differences offer an important observational test, as measurements of galaxy and cluster shapes from weak lensing or X-ray isophotes can constrain deviations from triaxiality, thereby providing indirect probes of the dark matter microphysics \citep{newman13b, duffy22, harvey24}, as we have shown in the two last sections.

\begin{table*}
\caption{\label{tab_par} Best fit parameters describing the relation of $b/a$ as a function of $c/a$ as in Eq.~(\ref{eqthirdorder}).}
\centering    
\begin{tabular}{l|cc|cc}
\hline \hline
 $ $ & DMO ($p_0$, $p_1$) &  $\chi^2_{\nu,DMO}$ & FP ($p_0$, $p_1$) & $\chi^2_{\nu,FP}$ \\  \hline \hline
$R_{200}$ & $0.365\pm 0.021$, $5.125 \pm 0.490$ & $0.021$ & $0.373\pm0.015$, $5.511\pm0.371$ & $0.103$ \\
$R_{200}/2$ & $0.377\pm0.015$, $5.173 \pm 0.444$ & $0.020$ & $0.383\pm0.012$, $6.377\pm0.343$ & $0.373$ \\
$R_{200}/10$ & $0.392\pm0.013$, $5.724 \pm 0.308$ & $0.626$ & $0.403\pm0.010$, $6.360\pm0.370$ & $0.188$ \\ \hline \hline
\end{tabular}
\end{table*}

\section{Summary and conclusions}
\label{sec_sum}
In this work, we have presented the first analysis of three-dimensional halo shapes, at $z=0$, in the AIDA-TNG suite, a set of cosmological hydrodynamical simulations designed to explore how alternative dark matter models -- Cold (CDM), Warm (WDM), and Self-Interacting (SIDM) -- affect the structure of halos in the presence of realistic baryonic physics. By comparing DMO and FP runs, we quantified how both microphysical interactions and baryonic feedback processes contribute to shaping the internal morphology and alignment of cosmic structures.

Our main results can be summarized as follows:

\begin{itemize}
    \item Halo shapes vary systematically with radius and component. The dark matter dominates and characterizes the overall morphology beyond $\sim R_{200}/2$, while baryonic condensation and feedback mainly reshape the inner regions. Gas is generally smoother and more oblate, while stars have a more triaxial distribution.
    
    \item Including baryonic physics makes halos significantly rounder than in the DMO case. The degree of sphericalization is largest in low- and intermediate-mass halos, where gas cooling and star formation deepen the potential well, while AGN feedback in massive systems prevents further condensation and maintains residual triaxiality. In these systems, feedback primarily prevents the inflow of fresh gas rather than erasing the pre-existing anisotropies.

    \item SIDM models produce rounder, more isotropic inner halos than CDM due to momentum transfer between dark matter particles, while WDM halos are both less concentrated and slightly more spherical. 
    The inclusion of baryonic effects reduces but does not eliminate these morphological differences, suggesting that the combined effect of baryonic feedback and dark matter self-interactions shapes the inner halo structure in a non-trivial manner.

    \item The orientation of stars, gas, and dark matter differs systematically across DM models and halo masses. In CDM, stars are typically aligned within $15$–$30^{\circ}$ with the dark matter, while SIDM induces larger misalignment angles, especially in massive halos where isotropization weakens the coupling between components. Gas orientations are more sensitive to feedback, becoming uncorrelated with the potential in strongly AGN-dominated systems, and sensitive to the considered dark matter model. 
    In addition, we find median misalignment between the BCG and host dark matter halo -- over all halo masses --  of 31$^{\circ}$, 38$^{\circ}$, 34$^{\circ}$, and 33$^{\circ}$  for CDM, SIDM1, vSIDM, and WDM3, respectively. CDM halos are slightly more aligned, while SIDM models show larger offsets due to isotropization of the dark matter potential. The misalignment decreases with stellar and halo mass, providing a key diagnostic for interpreting weak-lensing ellipticities and intrinsic alignments in upcoming surveys.

    \item The minor-to-major axis ratio $c/a$ decreases with increasing halo mass, confirming that more massive halos are more triaxial. Baryonic physics results in rounder configurations in which $c/a$ is larger by a mass-independent factor, indicating a nearly mass-independent sphericalization effect. Our results reproduce and extend previous findings \citep[e.g.:][]{despali14,bonamigo15} to alternative dark matter scenarios.  

    \item The relation between $b/a$ and $c/a$ follows a tight sequence, with halos being preferentially prolate in the DMO runs and more oblate in the FP case. This transition toward axisymmetry highlights the isotropizing influence of baryonic processes and, to a lesser degree, of dark matter self-interactions. In the FP runs, the central regions tend to be progressively more oblate or spherical; the opposite is true for halos in the dark matter-only simulations. 
\end{itemize}

Overall, our study demonstrates that baryonic feedback remains the dominant driver of halo roundness and alignment in the AIDA-TNG simulations, while alternative dark matter physics introduces additional, model-dependent modifications that are most pronounced in the inner regions. The combination of these effects yields distinctive, testable predictions for the three-dimensional morphology of galaxies and clusters.  In a recent series of papers \cite{velmani24, velmani25}, the authors have explored physically motivated parameterizations of baryonic feedback and its back-reaction on the dark matter distribution.  We hope our work stimulates further studies along these lines, both in extending beyond the spherical approximation and in considering alternative DM models.  

In future work, we will extend our analysis to higher redshifts, exploring the redshift evolution of halo shapes, and link the intrinsic 3D geometry to 2D observable quantities such as weak-lensing ellipticities, X-ray isophotes, and satellite galaxy anisotropies. These comparisons will provide a powerful route to constrain both baryonic feedback efficiency and the fundamental nature of dark matter.

\begin{acknowledgements}
LM and CG acknowledge the financial contribution from the PRIN-MUR
2022 20227RNLY3 grant 'The concordance cosmological model:
stress-tests with galaxy clusters' supported by Next Generation EU, and
from the grant ASI n. 2024-10-HH.0 “Attività scientifiche per la
missione Euclid – fase E”.  \\ GC thanks the support from INAF theory
Grant 2022: Illuminating Dark Matter using Weak Lensing by Cluster
Satellites. \\ GD acknowledges the funding by the European Union -
NextGenerationEU, in the framework of the HPC project – “National
Centre for HPC, Big Data and Quantum Computing” (PNRR - M4C2 - I1.4 -
CN00000013 – CUP J33C22001170001).  \\
RKS is grateful to the ISA Bologna for support and to the members of DIFA/INAF Bologna for their hospitality in the fall of 2025.\\
We acknowledge the EuroHPC Joint Undertaking for awarding the AIDA-TNG project access to the EuroHPC supercomputer LUMI, hosted by CSC (Finland) and the LUMI consortium through a EuroHPC Extreme Scale Access call. GD acknowledges ISCRA and ICSC for awarding her access to the LEONARDO supercomputer, owned by the EuroHPC Joint Undertaking, hosted by CINECA (Italy).

\end{acknowledgements}
\bibliography{aa5840025}
\end{document}